\documentclass[submit]{smj}

\usepackage{graphicx,setspace,lscape,longtable}
\usepackage{mathrsfs,amsthm}
\usepackage{natbib,epsfig,pdfpages}
\usepackage{rotating}
\usepackage{float}
\usepackage{natbib}
\usepackage{bm}
\usepackage[normalem]{ulem}
\usepackage{CJK}
\usepackage{subfigure}
\usepackage{array,booktabs}
\usepackage[titletoc]{appendix}
\usepackage{algorithm}
\usepackage{algpseudocode}

\bibpunct{(}{)}{;}{a}{,}{,}

\def\beq{\begin{equation}}
\def\eeq{\end{equation}}
\def\beqr{\begin{eqnarray}}
\def\eeqr{\end{eqnarray}}
\def\beqrs{\begin{eqnarray*}}
\def\eeqrs{\end{eqnarray*}}
\def\bc{\begin{center}}
\def\ec{\end{center}}

\Author{Huiwen Wang\Affil{1,2},
        Tingting Huang\Affil{1,3},
        and Shanshan Wang\Affil{1,3}
}
\AuthorRunning{Shanshan Wang \textrm{et al.}}

\Affiliations{

\item School of Economics and Management,
      Beihang University,
      Beijing,
      China
\item Beijing Advanced Innovation Center for Big Data and Brain Computing,
      Beihang University,
      Beijing,
      China
\item Beijing Key Laboratory of Emergence Support Simulation Technologies for City Operations,
      Beijing,
      China

}   

\CorrAddress{Shanshan Wang,
             School of Economics and Management,
             Beihang University,
             Xueyuan Road No. 37,
             Haidian District,
             Beijing,
             China}
\CorrEmail{sswang@buaa.edu.cn}
\CorrPhone{(+86)\;010\;82339337}
\CorrFax{(+86)\;010\;82328037}

\Title{A Flexible Spatial Autoregressive Modelling Framework for Mixed Covariates of Multiple Data Types}

\Abstract{
Mixed spatial autoregressive (SAR) models with numerical covariates have been well studied. However, as non-numerical data, such as functional data and compositional data, receive substantial amounts of attention and are applied to economics, medicine and meteorology, it becomes necessary to develop flexible SAR models with multiple data types. In this article, we integrate three types of covariates, functional, compositional and numerical, in an SAR model. The new model has the merits of classical functional linear models and compositional linear models with scalar responses. Moreover, we develop an estimation method for the proposed model, which is based on functional principal component analysis (FPCA), the isometric logratio (ilr) transformation and the maximum likelihood estimation method. Monte Carlo experiments demonstrate the effectiveness of the estimators. A real dataset is also used to illustrate the utility of the proposed model.
}


\begin{document}

\maketitle

\section{Introduction} \label{Sec1}
The mixed spatial autoregressive (SAR) model (\cite{Lesage2009Introduction}) gives a clear explanation of spatial spillover effects and influences from friends in social network. It has become popular to model activities in regional economies, social networks and spatial geography (\cite{Case1991Spatial,Topa2001Social,Olubusoye2016MODELLING}). Using a spatial weight matrix and a spatial lag parameter, the SAR model incorporates the network structure into a classical linear model. This is written in matrix form as
\beq \label{mod 1}
\bm{y}=\rho \bm{W}\bm{y}+\bm{x\beta}+\bm{\epsilon},
\eeq
where $\bm{y}$ is an $n$-dimensional dependent variable, $\bm{x}$ is an $n\times p$ matrix of regressors, $\rho$ is a scalar parameter, $\bm{\beta}$ is a $p$-dimensional slope to be estimated, and $\bm{\epsilon}$ is an $n$-dimensional vector of i.i.d disturbances following multiple normal distributions with zero mean and finite variances. Here, $W$ is a pre-defined spatial weight matrix built according to peer relations, geographic locations, or economic indicators (\cite{Case1993Budget}). Many estimation methods have been developed to obtain the parameters of model (\ref{mod 1}), including the maximum likelihood estimation method (MLE \cite{Keith1975Estimation,Lee2004Asymptotic}), generalized moment estimator (GMM \cite{Kelejian1999A,Lee2007GMM}) and Markov Chain Monte Carlo method (MCMC \cite{Lesage2009Introduction}).

Variations of the SAR model have also been present to handle real problems. Regarding the association between predictors and responses, \cite{Su2010Profile} presents partially linear spatial autoregressive models; \cite{SUN2018359} put forwards functional-coefficient spatial autoregressive panel data models. And to accommodate inconsistency of network effects for different locations, \cite{Dou2016Generalized} propose a spatio-temporal model with unknown diagonal coefficients. Banded spatio-temporal autoregressions (\cite{Gao2018Banded}) have also been presented to solve the problem whereby the spatial matrix $W$ is subjectively defined by merging two items $\rho$ and $W$ into an unknown matrix. The predictors of these models are all numerical. We find little literatures considering SAR models with non-numerical covariates.

However, with the development of memory technology, data of various types are being collected. Among them, complex data, including functional data, compositional data and symbolic data, have been widely used in the field of economics, meteorology, geochemistry and biology (\cite{RamsaySilverman-506,Lancet2010Principal,Wang2013Multiple}). Such data are endowed with special characteristics. For example, functional data are high dimensional,  and compositional data have sum-to-one constraints. It is natural to consider complex covariates in SAR models.

Moreover, it is often the case that more than one type of non-numerical predictor is involved in the regressions, especially when the data are gathered from different sources. For instance, we can collect relative humidity data (functional data) from a weather bureau and GDP (gross domestic product) structure data (compositional data) from statistical yearbooks to study factors that influence air pollution. Therefore, there is a need to build a new regression that addresses multiple types of covariates in the framework of the SAR model. In this article, we focus on three types of explanatory variables: functional, compositional and numerical.

\begin{figure}[!h]
  \begin{center}
  \label{Fig.1}
    \resizebox{5cm}{4cm}{\includegraphics{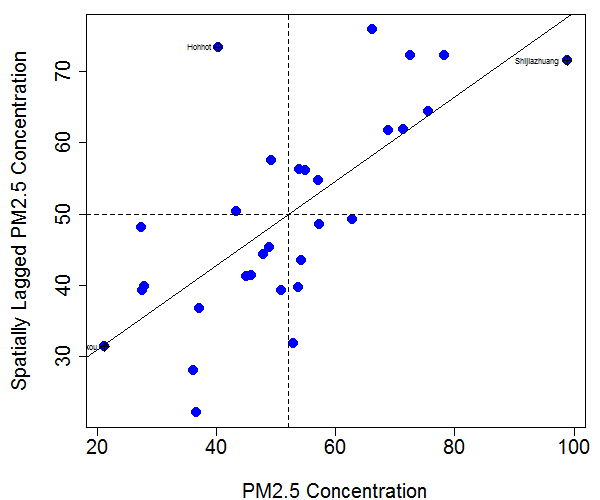}}
  \end{center}
  \caption{The Moran I scatter-plot of PM2.5 concentrations and spatially lagged PM2.5 concentrations.}
\end{figure}

Here, we use a real dataset to illustrate our motivation. The aim is to investigate how relative humidity, economic structure, GDP and GDP growth rate relate to annual mean PM2.5 (fine particulate matter smaller than 2.5 microns suspended in air) concentrations in $30$ major cities of China over 2016. The  PM2.5 concentration variable is the dependent variable. In preliminary data analysis, Moran's I statistic is adopted to examine whether spatial dependencies are present among the responses. Apparently, the value of Moran's I statistic is $0.53$, and the P value is less than $0.001$. These indicate that significant network dependencies exist and that an SAR model is adequate to fit the data. Moreover, it is obvious that the explanatory variables in the regression are of a mixed typed. To be clear, humidity data are monthly recorded functional data, the economic structures are compositional data consisting of three components (proportions of primary industry, secondary industry and tertiary industry), and GDP and GDP growth rate are numerical data. Thus, an SAR model with multiple types of covariates should be proposed. Figure \ref{Fig.1} shows Moran's I scatter-plot of PM2.5 concentrations and spatially lagged PM2.5 concentrations.

To absorb three data types in an SAR model, we should be concerned with operations in each data space. Fortunately, regressions of functional data and compositional data give us inspiration. When dependent and independent variables are scalar and functional, the classical functional linear model (\cite{TT2006Prediction,Hall2007Methodology}), which is built upon the functional inner product, can be utilised. And the compositional linear model relating compositional predictors to numerical responses (\cite{Hron2012Linear}) has also been presented based on the inner  product of the Aichison geometry. Borrowing techniques from these two types of models, we propose a flexible SAR model integrating functional, compositional and numerical covariates.

To the best of our knowledge, there has been minimal research considering mixed covariates. From a theoretical point of view, the proposed model combines the advantages of classical functional linear models, compositional linear models and mixed spatial autoregressive models, thereby providing a more flexible modelling framework for multiple types of covariates with spatial dependencies in the responses. We propose a maximum likelihood approach to estimate the regression parameter/function by incorporating the functional principal component analysis (FPCA) and isometric log-ratio (ilr) transformation to handle functional and compositional data, respectively. A Monte Carlo study is designed to examine the numerical performances of the estimators. In addition, we  use a real dataset to illustrate the utility of our model.

The article is organised as follows. In Section \ref{sec2}, we introduce some preliminaries for functional data and compositional data. The newly proposed model is presented in Section \ref{sec3}. In Section \ref{sec4}, the estimation method of the new model is explicitly elaborated upon. We conduct several numerical experiments in Section \ref{sec5} to evaluate the performance of the estimators. In addition, we employ the new model to analyse the PM2.5 concentration data in Section \ref{sec6}. Finally, the article is concluded with a discussion in Section \ref{sec7}.

\section{Preliminaries} \label{sec2}
In this section, we introduce the operations of inner products for compositional and functional data, which will help us understand the covariates of the proposed model.

We start with functional data. In practice, only the discrete values $x_{ij}$ of the sample curves $x_i(t)$ are recorded, where $j$ is the observation point. Thus, before analysing functional data, the first step is representing the raw data by curves. Commonly used methods contain basis expansion \cite{Ramsay2006Functional} and kernel functions \cite{Lu2006Nonparametric}.

When the underlying smooth functions $x_i(t)$ are obtained, we treat them as basic atoms. Specifically, we presume that $x(t)$ belongs to a $L^2$ space, which is composed of square integrable functions, i.e., $\int_0^1x^{2}(t)dt<\infty$. For simplicity, assume that all the functional data $x(t)$ are defined on the interval $[0, 1]$. The inner product of the functional data $x_1(t)$ and $x_2(t)$ is
$$\langle x_1(t), x_2(t)\rangle_{l^2}=\int_0^1 x_1(t)x_2(t)dt.$$
Here, the subscript $l^2$ denotes that the operation belongs to the $L^2$ space. We interpret $\langle x_1(t), x_2(t)\rangle_{l^2}$ as a projection of $x_1(t)$ on the function $x_2(t)$.

Then, we give definitions of compositional data. The sample space of compositional data is the simplex
$$S^D=\Big\{\bm{x}^D=(x_1^D, x_2^D, \dots, x_d^D)^\prime  \mid x_i^D >0, i=1,\dots,d; \sum_{i=1}^d x_i^D =1 \Big\},$$
where $\bm{x}^D$ is a $d$-part composition whose components are strictly positive and have a summation of $1$. The Aichison geometry is a Euclidean vector space built upon the simplex. Basic operations required for the Aichison geometry are perturbation, powering and inner product (\cite{Pawlowsky2015Modelling}).

Denote two $d$-part compositions as $\bm{x}^D=(x_1^D, x_2^D, \dots, x_d^D)^\prime$ and $\bm{y}^D=(y_1^D, y_2^D, \dots, y_d^D)^\prime$. The inner product of $\bm{x}^D, \bm{y}^D$ in the simplex is
$$\langle \bm{x}^D, \bm{y}^D \rangle_a=\sum_{i=1}^d \log \frac{x_i^D}{g_m(\bm{x}^D)} \log \frac{y_i^D}{g_m(\bm{y}^D)},$$
where $g_m(\bm{x}^D)$ is the geometric mean, i.e., $g_m(\bm{x}^D)=(\prod _{i=1}^{d} x_i^D)^{1/d}$, and the subscript $a$ indicates that the operation is in the simplex. According to the inner product, the norm of $\bm{x}^{D}$ can be evaluated by
$$\parallel \bm{x}^{D} \parallel_a=\sqrt{\langle \bm{x}^{D}, \bm{x}^{D} \rangle _a}=\sqrt{\sum_{i=1}^d \Big(\log \frac{x^D_i}{g_m(\bm{x}^{D})}\Big)^2}.$$

Note that the inner product of the compositions and the inner product of the functionals are real values. This make it possible to form regressions with multiple types of data.

\section{The new model}\label{sec3}
First, we introduce the network structure assumptions under which the new model is constructed. Following article \cite{Jenish2009Central}, we presume that the spatial process we aim to model is located on a (possibly) unevenly spaced lattice $L\subseteq R^n,~n\geq1$, and all elements on $L$ are endowed with positions. To ensure that all elements are separated, the distance $d(u,v)$ between any two elements $u$ and $v$ on $L$ should be greater than $0$. Here, $d(u,v)$ is crucial for establishing the weight matrix $W$.

Second, we formulate the proposed model. There are $n$ observations $\{y_i,x_i(t),\bm{x}_i^D,x_{i}\}_{i=1}^n$ from lattice $L$. Here $\{x_i(t)\}_{i=1}^n$, $\{\bm{x}_i^D=(x_{i1}^D,\dots,x_{id}^D)^\prime \}_{i=1}^n$ and $\{x_{i}\}_{i=1}^n$ are functional, compositional and numerical data, respectively. Denote $\bm{y}=(y_1,y_2,\dots,y_n)^\prime,~\bm{x}(t)=\big(x_1(t),x_2(t),\dots,x_n(t)\big)^\prime,~\bm{x}^D=(\bm{x}_1^D,\bm{x}_2^D,\dots,\bm{x}_n^D)^\prime
,~\bm{x}=(x_1,\dots,x_n)^\prime$. Then, $y_i$ are related to predictors by
\beq\label{equ1}
\bm{y}=\alpha\bm{\tau}_n+\rho \bm{Wy}+\langle\bm{x}(t),\beta(t)\rangle_{l^2}+\langle \bm{x}^{D},\bm{\beta}^{D}\rangle_a+\langle\bm{x},\beta\rangle_r+\bm{\epsilon}
\eeq
where $\alpha$ is an unknown constant; $\bm{\tau}_n$ is an $n$-dimensional column vector of ones; $\rho$ is the spatial lag parameter constrained on $(-1,1)$; $\bm{W}=(w_{ij})_{n\times n}$ is a prespecified spatial weight matrix whose diagonal elements are zero and whose sum of row elements is $1$; and $\bm{\epsilon}$ is an error term independent of $\bm{x}(t), ~\bm{x}^D, ~\bm{x}$ and follows a multivariate normal distribution, whose mean is $\bm{0}$ and covariance matrix is $\sigma^2 \bm{I}_n$, where $\bm{I}_n$ is an $n\times n$ identity matrix. $\beta(t),~\bm{\beta}^D,~\beta$ are coefficients to be estimated. Here, $\langle\cdot,\cdot\rangle_{r}$ denotes the inner product operation in the real vector space.

We interpret the spatial matrix $\bm{W}$ as a measure of the linkage strength for the spatial units on $L$, similar to \cite{Qu2015Estimating}. $\beta(t)$ is the best projection direction whereby $x_i(t)$ explains $y_i$ in model (\ref{equ1}). We also regard $\beta(t)$ as a weight function that assigns a particular coefficient $\beta(t_j)$ to $x_i(t_j)$ at any time $t_j,~t_j\in [0,1]$. Thus, $\beta(t)$ is a slope varying with $t$, which is more flexible compared to using a constant coefficient.  $\bm{\beta}^D$ is an elastic coefficient as well. If $\bm{x}_i^D$ has an alteration in $\frac{\bm{\beta}^D}{\parallel \bm{\beta}^D \parallel_a}$,
i.e., $\bm{x}_i^D+\frac{\bm{\beta}^D}{\parallel \bm{\beta}^D \parallel_a}$, then $y_i$ increases in $1$. Because $\langle \bm{x}_i^D+\frac{\bm{\beta}^D}{\parallel \bm{\beta}^D \parallel_a}, \bm{\beta}^D\rangle=\langle \bm{x}_i^D, \bm{\beta}^D \rangle+1$.

Our new model is flexible, as it reduces to several classical linear models in special cases.
\begin{itemize}
  \item When $\rho=0$, our model degenerates into the following linear model without a network:
  $$\bm{y}=\alpha\bm{\tau}_n+\langle\bm{x}(t),\beta(t)\rangle_{l^2}+\langle \bm{x}^{D},\bm{\beta}^{D}\rangle_a+\langle\bm{x},\beta\rangle_r+\bm{\epsilon}.$$
  \item When $\rho=0$ and the compositional predictor $\bm{x}^D$ is not present in the regression, our model is a functional linear model with numerical regressors $$\bm{y}=\alpha\bm{\tau}_n+\int_0^1\bm{x}(t)\beta(t)dt+\bm{x}\beta+\bm{\epsilon}.$$
  \item When $\rho=0$, $\bm{x}(t)$ is independent of $t$ and  numerical covariates are not present, our model becomes the compositional linear model $$\bm{y}=\langle \bm{x}^{D},\bm{\beta}^{D}\rangle_a+\bm{\epsilon}.$$
  \item When the compositional and functional regressors are not present, our model is the MSAR model (\ref{mod 1}).
\end{itemize}
Model (\ref{equ1}) can also be written as follows:
$$\bm{y}=\big(\bm{I}_n-\rho \bm{W}\big)^{-1}\big(\alpha\bm{\tau}_n+\langle \bm{x}(t),\beta(t)\rangle_{l^2}+\langle \bm{x}^{D},\bm{\beta}^{D}\rangle_a+\langle\bm{x},\beta\rangle_r+\bm{\epsilon}\big).$$
 It can be seen that the new errors $(\bm{I}_n-\rho \bm{W})^{-1}\bm{\epsilon}$ are not independent. In this situation, we use the maximum likelihood estimation method to obtain the parameters.

\section{Estimation method} \label{sec4}
In this section, we elaborate on the procedure for obtaining the  estimators of our proposed model. Addressing functional and compositional data in regression is difficult but crucial . To transform the infinite dimensional functional predictor into manageable finite variables, we employ FPCA techniques. Regarding compositional data, the sum constraint $\sum_{i=1}^d x_i=1$ makes the covariates singular. Applying compositions directly to the regression models will produce various issues. To solve these problems, we utilise the ilr transformation to equivalently express the compositions by real vectors.

The estimation process includes three steps: The first step is presenting the functional regressor by a  functional principal basis through FPCA; the second step is applying the ilr transformation, which makes the compositional covariate
processible; and the last step is using the maximum likelihood estimation method to estimate the expression obtained from Step $2$.

\subsection{Functional principal component analysis (FPCA)}\label{sec2-3}
We give a brief introduction of FPCA and then display how to address functional term in (\ref{equ1}).

In Section \ref{sec2}, we have assumed that the functional variable $X(t)$ is square integrable. Here, $X(t)$ is also random. Denote the covariance function of $X(t)$ by $K(s,t)$, i.e., $K(s,t)=\mbox{Cov}(X(t),X(s))$. Then, following Mercer's theorem, $K(s,t)$ admits the spectral decomposition
$$K(s,t)=\sum_{j=1}^{\infty}k_j \varphi_j (s) \varphi_j(t),~~~~k_1>k_2>\cdots>0,$$
where $k_j$ and $\varphi_j(t)$ are the corresponding eigenvalues and eigenfunctions, respectively. In addition, based on Karhunen-Lo\`{e}ve representation, $X(t)$ can be expanded as
$$X(t)=\sum_{j=1}^{\infty}a_j \varphi_j (t),$$
where $a_j$ are independent stochastic variables, with mean  $0$ and variance  $E(a_j^{2})=k_{j}$. Here, $a_j=\langle X(t),\varphi_j(t)\rangle_{l^2}=\int_{0}^{1} X(t)\varphi_j(t)dt$. Note that $\varphi_j(t)$ are a set of orthonormal bases in $L^2$. Thus, any square integrable functions can be expanded in $\varphi_j(t)$.

Because $\varphi_j(t)$ are theoretical, the empirical version of $\varphi_j(t)$ should be given. In practice, there are $n$ observations of the random function  $X(t)$, denoted by $\bm{x}(t)=\big(x_1(t),x_2 (t),\cdots,x_n (t)\big)^\prime$. Then, the empirical expression of $K(s,t)$ is
$$\hat{K}(s,t)=\frac{1}{n} \sum_{i=1}^{n}x_i (s) x_i (t)- \bar{x}(s) \bar{x}(t)$$,
where $\bar{x}(t)=\frac{1}{n} \sum_{i=1}^{n}x_i (t)$. We can also decompose $\hat K(s,t)$ into a sum of eigenfunctions,
$$\hat K(s,t)=\sum_{j=1}^{n}\hat k_j \hat \varphi_j (s) \hat\varphi_j(t),$$
where $\hat k_j$ and $\hat\varphi_j(t)$ are the estimators of $k_j$ and $\varphi_j(t)$, respectively. Because $\{\hat{\varphi}_j(t)\}_{j=1}^n$ is an orthonormal functional basis, the $i$th observation $x_i(t)$ can be expressed as $$x_i (t)=\sum_{j=1}^n\hat{a}_{ij} \hat{\varphi}_j (t),$$
 where $\hat a_{ij}=\langle x_i(t),\hat{\varphi}_j(t)\rangle_{l^2}$.

 For an unknown slope function $\beta(t)$, it has the following decomposed expression as well:
$$\beta(t)=\sum_{j=1}^n b_j \hat{\varphi}_j(t),$$ where $b_j=\langle\beta(t), \hat\varphi_j(t)\rangle_{l^2}$. Here, $b_j$ are some unknown real values that need to be determined. We treat them as known at the moment. It is easy to find that the inner product of $x_i(t)$ and $\beta(t)$ can be evaluated by $\hat a_{ij}$ and $b_j$, and we have
$$\langle x_{i}(t), \beta(t)\rangle_{l^2}=\sum_{j=1}^n\hat a_{ij}b_j.$$ Therefore, $\langle \bm{x}(t), \beta(t)\rangle_{l^2}$ in model (\ref{equ1}) can be replaced by $\sum_{j=1}^n\hat {\bm{a}}_{j}b_j$:
\beq \label{tran.1}
\bm{y}=\alpha \bm{\tau}_n+\rho \bm{Wy}+\sum_{j=1}^{n}\bm{\hat{a}_{j}} b_j+\langle \bm{x}^D, \bm{\beta}^D\rangle_{a}+\langle\bm{x},\beta\rangle_r+\bm{\epsilon},
\eeq
where $\hat {\bm{a}}_{j}=(a_{1j},\dots,a_{nj})^\prime.$

Note that in reality, $x_i(t)$ can be well approximated by the first $m$ principal components.  The  percentages of variances explained (PVE) criterion for covariates is often used to choose the truncation parameter $m$. If we set the PVE to  $z$, the parameter $m$ is subject to $\underset{l}{\mbox{min}}~\{(\sum_{j=1}^{l}\hat{k}_{j})/(\sum_{j=1}^{n}\hat{k}_j)\geq z\}$. In our numerical experiments, $m$ is selected using this technique. Therefore, we approximate model (\ref{tran.1}) as
\beq\label{equ5}
\bm{y}\approx \alpha \bm{\tau}_n+\rho \bm{Wy}+\sum_{j=1}^{m}\bm{\hat{a}_{j}} b_j+\langle \bm{x}^D, \bm{\beta}^D\rangle_{a}+\langle\bm{x},\beta\rangle_r+\bm{\epsilon}
\eeq

\subsection{Isometric log-ratio (ilr) transformation}\label{sec2-2}
The main point of the ilr transformation is in representing a $d$-part dependent composition $\bm{x}^D=(x_1^D, x_2^D, \dots, x_d^D)^{\prime}$ by a $d-1$-dimensional independent vector $\bm{\xi}=(\xi_1, \xi_2, \dots, \xi_{d-1})^{\prime}$. In the following, we first introduce the theory of the ilr transformation and then present the detailed operation of this mapping.

We assume that $\{\bm{e}_k^D \}_{k=1}^{d-1}, \bm{e}_k^D=(e_{k1}^D, e_{k2}^D, \dots, e_{kd}^D)^{\prime}$ is a set of orthonormal bases in the Aichison geometry (\cite{Pawlowsky-Glahn}). A $d$-part composition $\bm{x}^D$ can then be expanded as
$$\bm{x}^D=\langle \bm{x}^D, \bm{e}_1^D \rangle_a \odot\bm{e}_1^D \oplus \langle \bm{x}^D, \bm{e}_2^D \rangle_a \odot \bm{e}_2^D \oplus \dots \oplus \langle \bm{x}^D, \bm{e}_{d-1}^D \rangle_a \odot \bm{e}_{d-1}^D.$$
where $\odot, \oplus$ are  powering and perturbation operations  in the Aichison geometry (\cite{Pawlowsky-Glahn}).
The ilr transformation of $\bm{x}^D$ is the coefficient of the basis $\{\bm{e}_k^D \}_{k=1}^{d-1}$,
$$\bm{\xi}=ilr(\bm{x}^D)=(\langle \bm{x}^D, \bm{e}_1^D \rangle_a, \langle \bm{x}^D, \bm{e}_2^D \rangle_a, \dots, \langle \bm{x}^D, \bm{e}_{d-1}^D \rangle_a)^{\prime}.$$
Note that $\bm{\xi}$ varies with the choice of basis $\{\bm{e}_k^D \}_{k=1}^{d-1}$. In this article, we use a sequential binary partitioning method to construct the orthonormal basis (\cite{Egozcue-Paw}). The ilr coordinates under this basis are
\beq\label{ilr 1}
\xi_i=\sqrt{\frac{d-i}{d-i+1}}\ln\frac{x_i^D}{\sqrt[d-i]{\prod_{j=i+1}^Dx_j^D}},~~~~i=1,2,\dots,d-1.
\eeq
Then, the inverse of this ilr transformation $ilr^{-1}$, which obtains the original compositions from the coordinates, is given by
$$x_1^D=exp\big(\frac{\sqrt{d-1}}{\sqrt{d}}\xi_1\big),$$
$$x_d^D=exp\big(-\sum_{j=1}^{d-1}\frac{1}{\sqrt{(d-j+1)(d-j)}}\xi_j \big),$$ and
$$x_i^D=exp\big(-\sum_{j=1}^{i-1}\frac{1}{\sqrt{(d-j+1)(d-j)}}\xi_j+\frac{\sqrt{d-i}}{\sqrt{d-i+1}}\xi_j\big),~~~~j=2,\dots,d-1.$$
We can know that the ilr transformation is reversible.

Now, there are $n$ compositional observations $\{\bm{x}_i^D\}_{i=1}^n$, $\bm{x}_i^D=(x_{i1}^D,\dots,x_{id}^D)^\prime$. We can transform them into ilr variable samples $\{\bm{\xi_i}\}_{i=1}^n$, $\bm{\xi}_i=(\xi_{i1},\dots,\xi_{i(d-1)})^\prime$ using conversion (\ref{ilr 1}). The coefficient $\bm{\beta}^D$ can be disposed of in the same manner,
$$ilr(\bm{\beta}^D)=\bm{\theta}=(\theta_1,\dots,\theta_{d-1})^\prime,$$
where $\bm{\theta}$ is unknown. We regard $\bm{\theta}$ as a parameter. Because the ilr transformation keeps the inner product unchanged, that is,
$\langle \bm{x}_i^D, \bm{\beta}^D\rangle_{a}=\langle \bm{\xi_i}, \bm{\theta}\rangle_r$. We substitute $\langle \bm{x}^D, \bm{\beta}^D\rangle_{a}$ by $\bm{\xi}\bm{\theta}$ in model (\ref{equ1}), where  $\bm{\xi}=(\bm{\xi}_1,\dots,\bm{\xi}_n)^\prime$. The new expression is
\beq\label{equ4}
\bm{y}\approx \alpha \bm{\tau}_n+\rho \bm{Wy}+\sum_{j=1}^{m}\bm{\hat{a}_{j}} b_j+\bm{\xi}\bm{\theta}+\bm{x}\beta+\bm{\epsilon}.
\eeq

\subsection{Maximum likelihood estimation method (MLE)}
In this subsection, the maximum likelihood estimation method is used to estimate the truncated model (\ref{equ4}).

Denote $\bm{A}= ( \hat{a}_{ij})_{n\times m}$, $\bm{b}=(b_1, b_2, \cdots, b_m)^\prime$, $\bm{Z}=(\bm{\tau}_n,\bm{A},\bm{\xi},\bm{x})$, and $\bm{\delta}=(\alpha,\bm{b},\bm{\theta},\beta)^{\prime}$. Model (\ref{equ4}) has the simple form
\beq\label{equ6}
\bm{y}\approx\rho \bm{Wy}+\bm{Z\delta}+\bm{\epsilon}.
\eeq
Clearly, expression (\ref{equ6}) is similar to an SAR model. Therefore, it is straightforward to use MLE, which is a popular estimation method for the SAR model, to obtain estimators in (\ref{equ6}).

Because the error term $\epsilon$ follows a multivariate normal distribution, the distribution of $\bm{y}$ can be derived. Then, the log-likelihood function of $\bm{y}$ is
\beq\label{log-lik}
\ln L(\rho, \bm{\delta}, \sigma^2)=-\frac{n}{2}\ln(2\pi\sigma^2)+\ln|\bm{I}_n-\rho \bm{W}|-\frac{\bm{e}^\prime \bm{e}}{2\sigma ^2},
\eeq
where $\bm{e}=\bm{y}- \rho \bm{Wy}- \bm{Z\delta}$. Notice that there are three variables in (\ref{log-lik}); thus, it is difficult to obtain the maximum value. However, if the estimator of $\rho$ is obtained, the estimators of $\bm{\delta}$ and $\sigma^2$ can be accordingly derived as follows:
\begin{eqnarray}
  \bm{\hat{\delta}}(\rho) &=& (\bm{Z}^\prime \bm{Z})^{-1}\bm{Z}^\prime(\bm{\bm{I}_n}-\rho \bm{W})\bm{y},\label{deltahat} \\
  \hat{\sigma}^2 (\rho) &=& (\bm{y}-\rho \bm{Wy}-\bm{Z}\bm{\hat{\delta}}(\rho))^{\prime}(\bm{y}-\rho \bm{Wy}-\bm{Z}\bm{\hat{\delta}}(\rho))/n.\label{sigmahat}
\end{eqnarray}
We can substitute $\bm{\delta}$ and $\sigma^2$ in (\ref{log-lik}) using (\ref{deltahat}) and (\ref{sigmahat}); then, the maximum of  function (\ref{log-lik}) is evaluated as follows:
\beq\label{hat.rho}
\hat{\rho}=\mbox{arg}\max_\rho \Big\{-\frac{n}{2}\ln(\hat{\sigma}^2(\rho))+\ln|\bm{I}_n-\rho \bm{W}|\Big\}.
\eeq
Optimisation methods, such as Newton's method, can be used to obtain the numerical solution.

As long as $\hat{\rho}$ can be obtained, $\hat{\bm{\delta}}$ and $\hat{\sigma}^2$ can be derived. Then, the estimators of $\beta(t), ~\bm{\beta}^D$ are evaluated by
\begin{eqnarray}
\hat{\beta}(t)=\sum_{j=1}^{m}\hat{b}_j\hat{\varphi}_j(t) \label{func.beta} \\
\hat{\bm{\beta}}^D=ilr^{-1}(\hat{\bm{\theta}}) \label{com.beta}.
\end{eqnarray}

We summarise the  estimation procedure as follows:
\vspace{3mm}
\begin{algorithm}[!h]
\begin{spacing}{1.0}
  \caption{Main steps of the estimation procedure}
  \begin{algorithmic}[1]
    \State Transform the functional predictor and the slope function into numerical data. In this step, we use functional principal basis to expand the functional term. Thus, we obtain expression (\ref{equ5}).
    \State Express the compositional covariate and the compositional coefficient with real vectors. In this step, we use the ilr transformation to equivalently represent compositions in ilr coordinates. The transformed model can be easily processed, as shown in (\ref{equ4}).
    \State Estimate the unknown parameters of the new transformed model obtained from Step 2. Because the new form of our model is similar to a MSAR model, the maximum likelihood estimation method is used to obtain the estimators. We estimate the spatial lag parameter $\rho$, coefficient $\bm{\delta}=(\alpha, \bm{b}, \bm{\theta})^\prime$, and variance $\sigma ^2$ from (\ref{hat.rho}), (\ref{deltahat}) and (\ref{sigmahat}), respectively. 
    \State Determine the estimators of $\beta(t)$ and $\bm{\beta}^D$ in our model. The functional coefficient is reconstructed in the FPC basis mentioned in Step 1 and with the coefficients $\bm{\hat{b}}$ estimated in Step 3, as shown in (\ref{func.beta}). The compositional slope is evaluated with the inverse of the ilr transformation and $\bm{\theta}$ from (\ref{com.beta}).
  \end{algorithmic}
  \end{spacing}
\end{algorithm}

\section{Numerical Experiments}\label{sec5}
To assess the performances of the estimation method, several experiments are conducted in this section. Specifically, three parameters that strongly influence the estimation results are discussed. In data preprocessing, the discrete values of the functional covariates are converted into curves using the Epanechnikov kernel. We set the PVE to  $70\%$ regarding the truncation parameter. All the studies are implemented in the R environment. In addition, we used the ``fda", ``spdep", ``compositions" and ``Compositional" packages.

\subsection{Critical parameters}\label{sec5-1}
We explain three parameters in this subsection: the spatial weight matrix $W$, spatial lag parameter $\rho$, and control parameter $\alpha$. They all impact performances of the proposed estimators, and can be interpreted as representing three aspects of our model. The spatial matrix reflects whether the network is closely connected. The value of $\rho$ indicates whether the spatial effects are strong in the regression. At last, the control parameter $\alpha$ relates to the characteristics of the functional predictor.

\begin{enumerate}
  \item \cite{Lee2004Asymptotic}  noted that the convergent rates of estimators depend on features of the spatial weight matrix $W$, when the maximum likelihood estimation method is used to get parameters of the SAR model. And in the scenarios where the units are affected by only a few neighbours, the estimators have a $\sqrt{n}$-rate of convergence. In our experiments, we adopt the commonly used rook matrix as spatial scenario, whose spatial weight matrix is sparse. Under such setting, $n$ individuals are randomly located on a grid of $R$ rows and $T$ columns, with each individual occupying a square. That is $n=R\times T$. Besides, regard two units as neighbours if they share a border. Then, the weight between two units is $1$ if they are neighbouring and $0$ otherwise.
 \item We have mentioned in Section \ref{sec3} that when $\rho$ equals $0$, the new model degenerates into a linear model with mixed covariates, which means that there are no spatial effects. To see how the strength of the network structure in the regression affects the performances of the estimators, three values of $\rho$ are considered: $0$, $0.4$ and $0.8$.
  \item \cite{Hall2007Methodology}  pointed out that the estimation accuracy of $\beta(t)$ of the classical functional linear model relies on the spaces between the eigenvalues of the sample covariance function,    when the estimation method is based on FPCA. To see whether the accuracies of the proposed estimators are related to the spaces between the eigenvalues, we design two cases, $\alpha=1.1$ and $\alpha=2$, following the parameter settings in \cite{Hall2007Methodology}. Note that when $\alpha=1.1$, the eigenvalues are well spaced, which is expected to result in more precise estimators.
\end{enumerate}

\subsection{Data generation process}\label{sec5-2}

The responses $\bm{y}=(y_1, y_2, \dots, y_n)^\prime$ are generated by
$$\bm{y}=(\bm{I}_n-\rho \bm{W})^{-1}\Big(\int_{0}^{1}\bm{x}(t)\beta(t)dt+\langle \bm{x}^{D},\bm{\beta}^{D}\rangle_a+\bm{x}\beta+0.5\bm{\epsilon}\Big),~~~~\bm{\epsilon}\sim N(\bm{0},\bm{I}_n).$$
$$\rho=\{0, 0.4, 0.8\},~~~~\bm{\beta}^D=\Big(\frac{4}{9}, \frac{2}{9}, \frac{1}{3}\Big)^\prime,~~~~\beta=1,~~~~x\sim N(1,0.5)$$
For the spatial scenarios, we set the parameters as follows:
 $$n=R\times T=\{10\times 15, 10\times 30, 30\times 30\}=\{150, 300, 900\}.$$

The functional predictor $x(t)$ takes the same form as that of FLM in \cite{Hall2007Methodology}. We generate the $x_i(t)$ independently from
$$x(t)=\sum_{j=1}^{50}a_jZ_j\phi _j(t),$$
where $a_j=(-1)^{j+1}j^{-\frac{\alpha}{2}}$, with $\alpha=1.1$ and $2$; $Z_j\sim U[\sqrt{3}, \sqrt{3}]$; and $\phi_j(t)=\sqrt{2}cos(j\pi t)$. And $\beta(t)$ is a linear combination of $\phi_j(t)$. It is generated as
$$\beta(t)=\sum_{j=1}^{50}b_j\phi_j(t),$$
where $b_j=4(-1)^{j+1}j^{-2},~j\geq2$ and $0.3$ when $j=1$.

With respect to the compositional covariate $\bm{x}^D$, let its expectation be $(\frac{1}{6}, \frac{1}{3}, \frac{1}{2})^\prime$. The covariance of $\bm{x}^D$ is expressed by the covariance matrix $\Sigma$ in the ilr coordinates of $\bm{x}^D$. We set
$$\Sigma=\left(
 \begin{array}{cc}
 2 & -1.5 \\
 -1.5 & 2 \\
\end{array}
\right)
.$$

There are $500$ repetitions for each setting. The performances of the estimators $\hat{\rho},~\hat{\beta}$ are evaluated with respect to mean bias and standard deviation. The performance of $\hat{\beta}(t)$ is evaluated in terms of the mean square error (MSE)
$$MSE(\hat{\beta}(t))=\frac{1}{100}\sum_{j=1}^{100}(\hat{\beta}(t_j)-\beta(t_j))^2,$$
where $t_j$ are $100$ equally spaced points on $[0,1]$. We assess the efficiency of $\hat{\bm{\beta}}^D$ through the mean bias of the components and simplicial standard deviation. Concretely, we denote the empirical mean of $\hat{\bm{\beta}}^D$ by $\overline{\bm{\beta}}^D=(\overline{\beta}_1^D, \overline{\beta}_2^D, \dots, \overline{\beta}_d^D)^\prime$, and the mean bias of the components is
$$bias(\beta_k^D)=E(\beta_k^D)-\beta_k^D=\overline{\beta}_k^D-\beta_k^D,~~~~k=1,\dots,d.$$
Here, $\overline{\bm{\beta}}^D=ilr^{-1}\big(\frac{1}{n}\sum_{i=1}^n\hat{\bm{\theta}}_i\big)$, where $\hat{\bm{\theta}}_i=(\hat{\theta}_{i1},\dots,\hat{\theta}_{id-1})^\prime$ is the ilr coordinate of the $i$th estimated $\bm{\beta}^D$.
In addition, The simplicial standard deviation of $\hat{\bm{\beta}}^D$ is
$$sstd(\hat{\bm{\beta}}^D)=\sqrt{\frac{totvar(\hat{\bm{\beta}}^D)}{d-1}},$$
where $totvar(\hat{\bm{\beta}}^D)$ is the total variance evaluated by
$totvar(\hat{\bm{\beta}}^D)=\sum_{j=1}^{d-1}var(\hat{\bm{\theta}}_j), \hat{\bm{\theta}}_j=(\hat{\theta}_{1j},\dots,\hat{\theta}_{nj})^\prime$.

\subsection{Results}\label{sec5-3}
\begin{table}[h!]
 \begin{spacing}{2.0}
\footnotesize
\caption{When $\alpha=1.1$, the mean biases and standard deviations (in bracket) of $\hat{\rho}$ and $\hat{\beta}$; the empirical average MSE and its standard deviation (in bracket) of $\hat{\beta}(t)$; and the components' mean biases and simplicial standard deviation (in bracket) of $\hat{\bm{\beta}}^D$.}
\vspace{-8mm}
  \setlength\tabcolsep{6pt}
  \renewcommand{\arraystretch}{1.5}
\label{Tab 1}
\bc
\begin{tabular}{cccccccccccc}
\\\specialrule{0.05em}{-1pt}{-1pt}
$\bm{\rho}$&$\bm{n}$& $\hat{\bm{\rho}}$& $\bm{\hat{\beta}(t)}$ & $\bm{\hat{\beta}}$ & $\bm{\hat{\beta}_1^D}$ & $\bm{\hat{\beta}_2^D}$ & $\bm{\hat{\beta}_3^D}$\\\specialrule{0.05em}{-1pt}{-1pt}
$\rho$=0&n=150& $\underset{(0.056)}{0.0011}$ & $\underset{(0.025)}{0.0700}$ & $\underset{(0.081)}{0.0005}$  & -0.0007 & $\underset{(0.004)}{0.0003}$  & 0.0004  \\\specialrule{0em}{-5pt}{-5pt}
&n=300& $\underset{(0.042)}{0.0001}$ & $\underset{(0.007)}{0.0204}$ & $\underset{(0.059)}{-0.0018}$  & 0.0004 & $\underset{(0.002)}{-0.0002}$  & -0.0002  \\\specialrule{0em}{-5pt}{-5pt}
&n=900& $\underset{(0.023)}{-0.0005}$ & $\underset{(0.001)}{0.0035}$ & $\underset{(0.035)}{0.0026}$  & 0.0006 & $\underset{(0.0007)}{-0.0001}$  & -0.0005  \\\specialrule{0em}{-5pt}{-5pt}
$\rho$=0.4&n=150& $\underset{(0.055)}{-0.008}$ & $\underset{(0.024)}{0.0677}$ & $\underset{(0.083)}{0.0010}$  & 0.0004 & $\underset{(0.004)}{0.0000 }$  & -0.0004  \\\specialrule{0em}{-5pt}{-5pt}
&n=300& $\underset{(0.037)}{-0.0047}$ & $\underset{(0.007)}{0.0202}$ & $\underset{(0.059)}{-0.0028}$  & 0.0014 & $\underset{(0.002)}{-0.0004}$  & -0.0010  \\\specialrule{0em}{-5pt}{-5pt}
&n=900& $\underset{(0.021)}{-0.0019}$ & $\underset{(0.001)}{0.0034}$ & $\underset{(0.033)}{0.0018}$  & -0.0001 & $\underset{(0.0006)}{0.0001}$  & 0.0000  \\\specialrule{0em}{-5pt}{-5pt}
$\rho$=0.8&n=150& $\underset{(0.030)}{-0.0075}$ & $\underset{(0.025)}{0.0695}$ & $\underset{(0.086)}{0.0026}$  & -0.0005 & $\underset{(0.004)}{-0.0001}$  & 0.0006  \\\specialrule{0em}{-5pt}{-5pt}
&n=300& $\underset{(0.019)}{-0.0043}$ & $\underset{(0.008)}{0.0211}$ & $\underset{(0.059)}{0.0014}$  & 0.0004 & $\underset{(0.002)}{-0.0001}$  & -0.0003  \\\specialrule{0em}{-5pt}{-5pt}
&n=900& $\underset{(0.012)}{-0.0020}$ & $\underset{(0.001)}{0.0034}$ & $\underset{(0.033)}{0.0003}$  & 0.0005 & $\underset{(0.0007)}{-0.0003}$  & -0.0002 \\\specialrule{0.05em}{-1pt}{-1pt}
\end{tabular}\ec
\end{spacing}
\end{table}

\begin{table}[h!]
 \begin{spacing}{2.0}
\footnotesize
\caption{When $\alpha=2$, the mean biases and standard deviations (in bracket) of $\hat{\rho}$ and $\hat{\beta}$; the empirical average MSE and its standard deviation (in bracket) of $\hat{\beta}(t)$; and the components' mean biases and simplicial standard deviation (in bracket) of $\hat{\bm{\beta}}^D$.}
\vspace{-8mm}
  \setlength\tabcolsep{6pt}
  \renewcommand{\arraystretch}{1.5}
\label{Tab 2}
\bc
\begin{tabular}{ccccccccccccccccccc}
\\\specialrule{0.05em}{-1pt}{-1pt}
$\bm{\rho}$&$\bm{n}$& $\hat{\bm{\rho}}$& $\bm{\hat{\beta}(t)}$ & $\bm{\hat{\beta}}$ & $\bm{\hat{\beta}_1^D}$ & $\bm{\hat{\beta}_2^D}$ & $\bm{\hat{\beta}_3^D}$\\\specialrule{0.05em}{-1pt}{-1pt}
$\rho$=0&n=150& $\underset{(0.064)}{-0.0061}$ & $\underset{(0.062)}{0.2441}$ & $\underset{(0.092)}{0.0041}$  & 0.0011 & $\underset{(0.004)}{-0.0001}$  & -0.0010  \\\specialrule{0em}{-5pt}{-5pt}
&n=300& $\underset{(0.044)}{0.0016}$ & $\underset{(0.023)}{0.1167}$ & $\underset{(0.060)}{0.0033}$  & 0.0001 & $\underset{(0.002)}{0.0001}$  & -0.0002  \\\specialrule{0em}{-5pt}{-5pt}
&n=900& $\underset{(0.026)}{-0.0016}$ & $\underset{(0.004)}{0.038}$ & $\underset{(0.035)}{-0.0005}$  & 0.0001 & $\underset{(0.0007)}{0.0000}$  & -0.0001  \\\specialrule{0em}{-5pt}{-5pt}
$\rho$=0.4&n=150& $\underset{(0.057)}{-0.0075}$ & $\underset{(0.059)}{0.2408}$ & $\underset{(0.085)}{0.0038}$  & 0.0005 & $\underset{(0.004)}{0.0001}$  & -0.0006  \\\specialrule{0em}{-5pt}{-5pt}
&n=300& $\underset{(0.039)}{-0.0050}$ & $\underset{(0.023)}{0.1167}$ & $\underset{(0.061)}{0.0027}$  & 0.0002 & $\underset{(0.002)}{-0.0001}$  & -0.0001  \\\specialrule{0em}{-5pt}{-5pt}
&n=900& $\underset{(0.022)}{-0.0016}$ & $\underset{(0.004)}{0.0377}$ & $\underset{(0.036)}{-0.0006}$  & -0.0005 & $\underset{(0.0007)}{0.0003}$  & 0.0002  \\\specialrule{0em}{-5pt}{-5pt}
$\rho$=0.8&n=150& $\underset{(0.035)}{-0.0114}$ & $\underset{(0.065)}{0.2428}$ & $\underset{(0.088)}{-0.0005}$  & -0.0001 & $\underset{(0.0046)}{-0.0003}$  & 0.0004  \\\specialrule{0em}{-5pt}{-5pt}
&n=300& $\underset{(0.022)}{-0.0041}$ & $\underset{(0.023)}{0.1157}$ & $\underset{(0.060)}{0.0007}$  & -0.0002 & $\underset{(0.002)}{-0.0002}$  & 0.0004  \\\specialrule{0em}{-5pt}{-5pt}
&n=900& $\underset{(0.013)}{-0.0010}$ & $\underset{(0.004)}{0.0379}$ & $\underset{(0.034)}{0.0009}$  & 0.0004 & $\underset{(0.0007)}{-0.0001}$  & -0.0003  \\\specialrule{0.05em}{-1pt}{-1pt}
\end{tabular}\ec
\end{spacing}
\end{table}

\begin{figure}[!h]
  \begin{center}
  \subfigure{
    \resizebox{4cm}{3.5cm}{\includegraphics{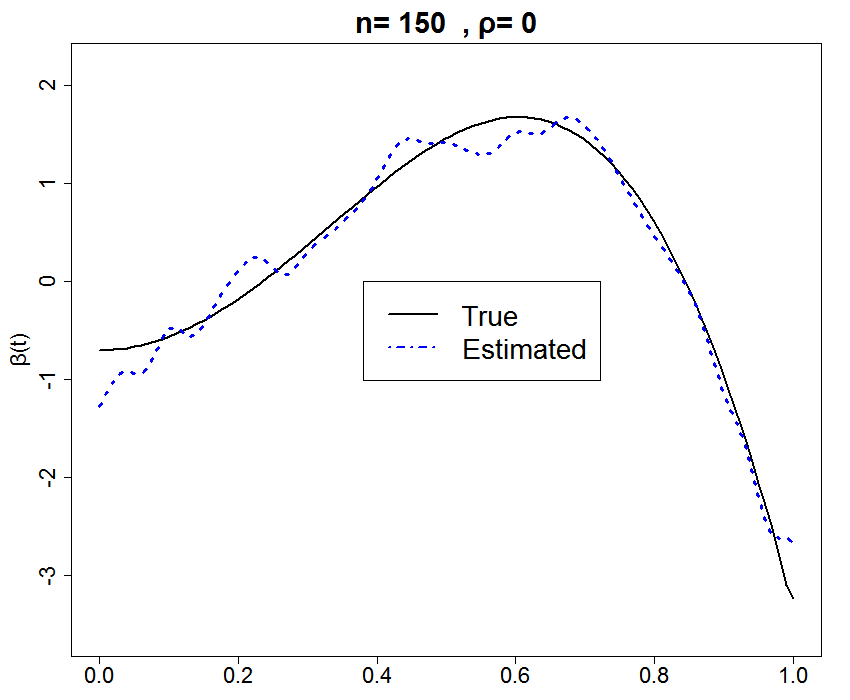}}}
    \subfigure{
    \resizebox{4cm}{3.5cm}{\includegraphics{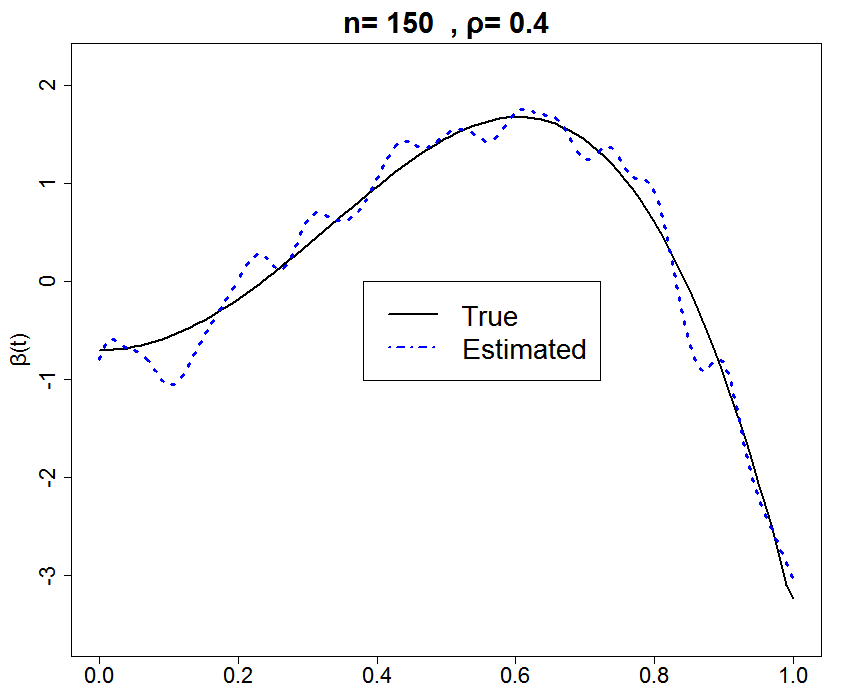}}}
    \subfigure{
    \resizebox{4cm}{3.5cm}{\includegraphics{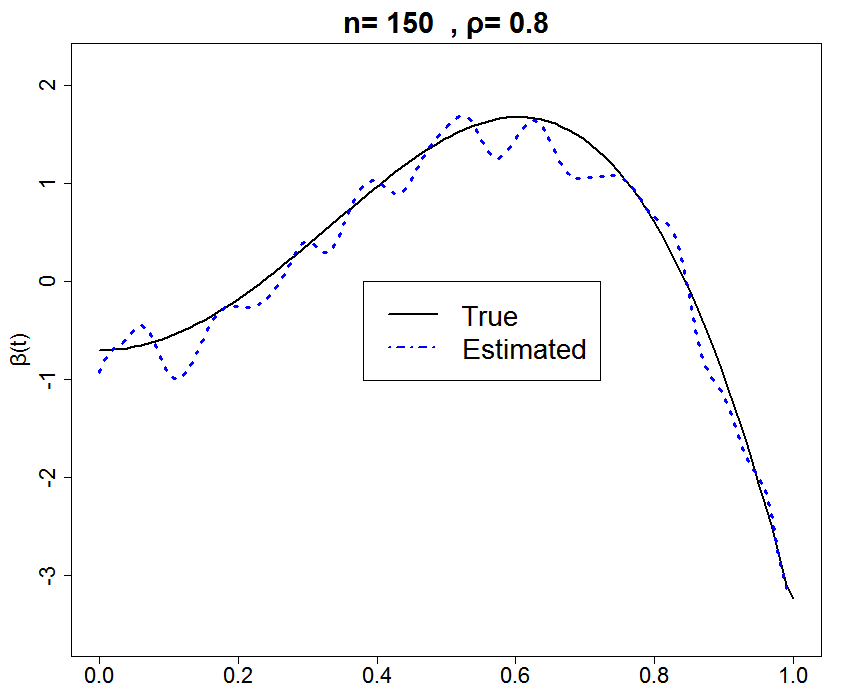}}}
      \subfigure{
    \resizebox{4cm}{3.5cm}{\includegraphics{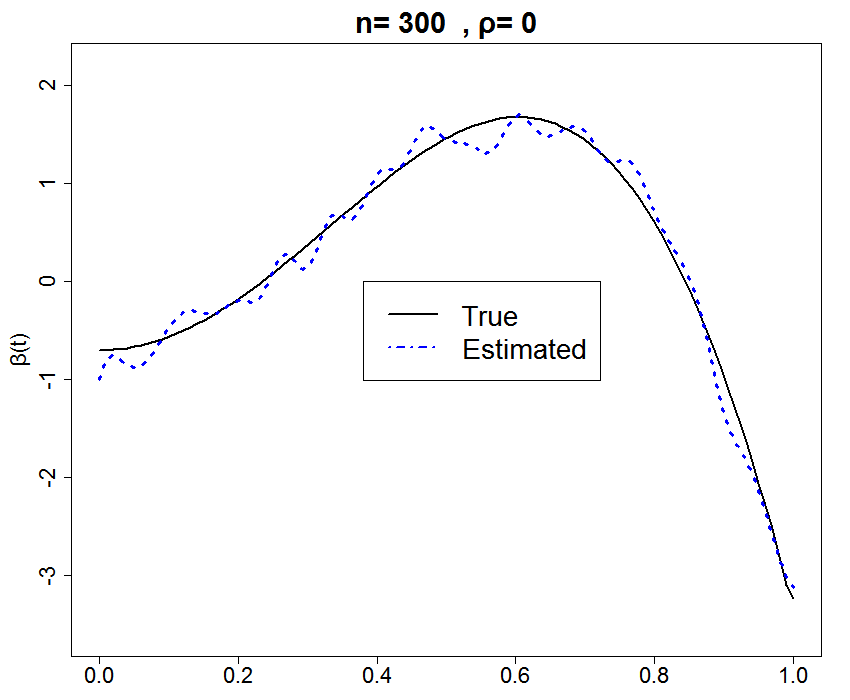}}}
    \subfigure{
    \resizebox{4cm}{3.5cm}{\includegraphics{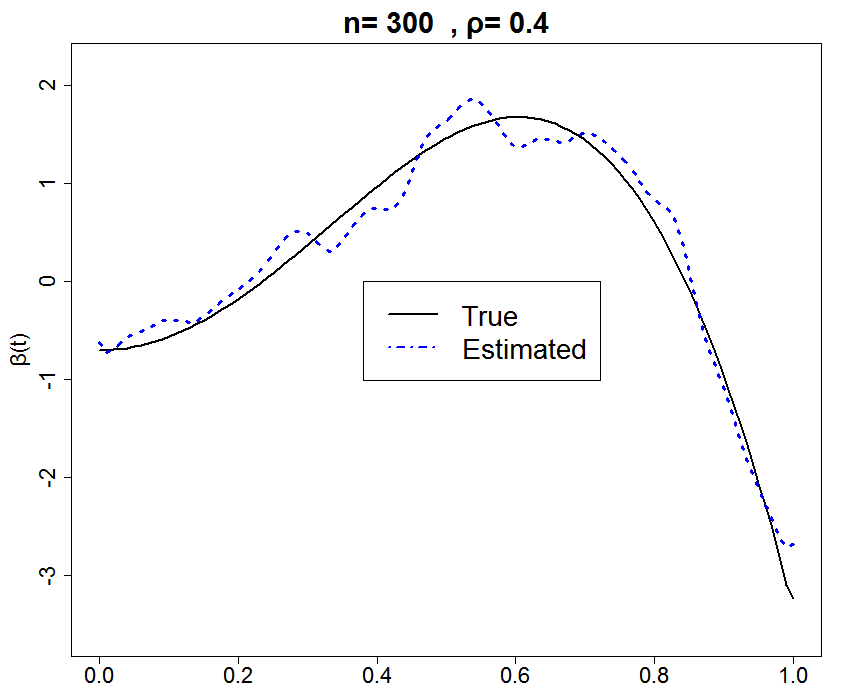}}}
    \subfigure{
    \resizebox{4cm}{3.5cm}{\includegraphics{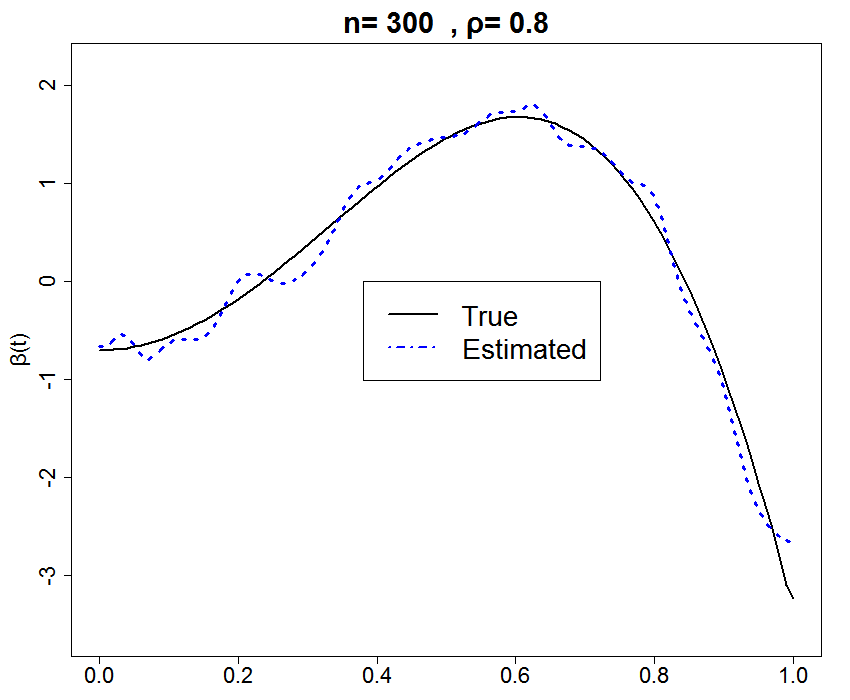}}}
    \subfigure{
    \resizebox{4cm}{3.5cm}{\includegraphics{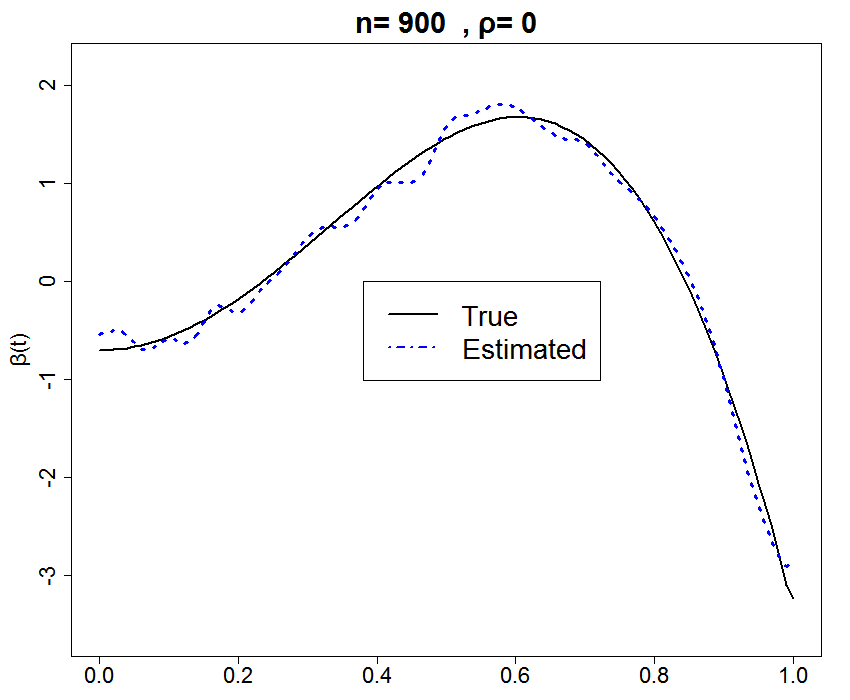}}}
    \subfigure{
    \resizebox{4cm}{3.5cm}{\includegraphics{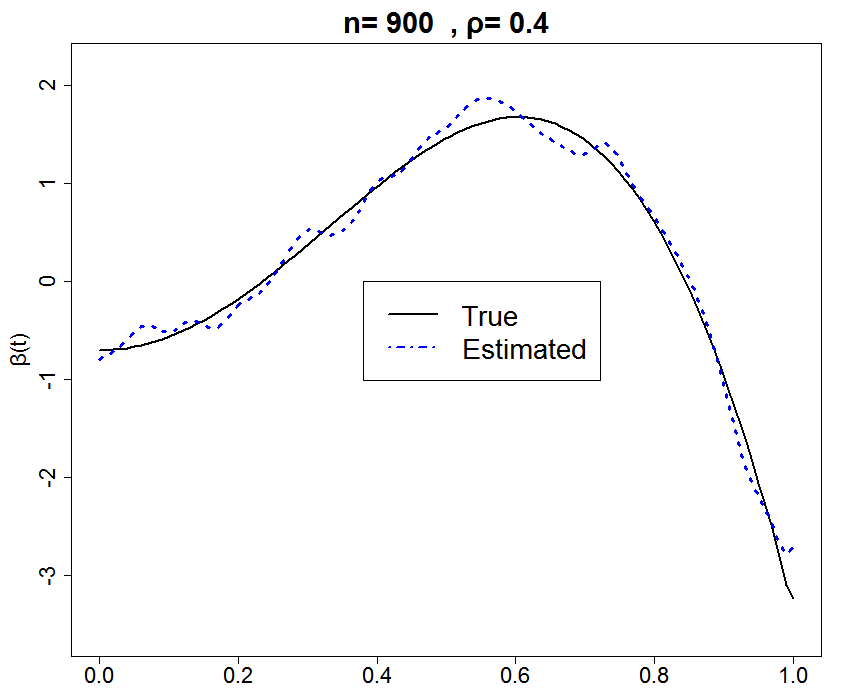}}}
    \subfigure{
    \resizebox{4cm}{3.5cm}{\includegraphics{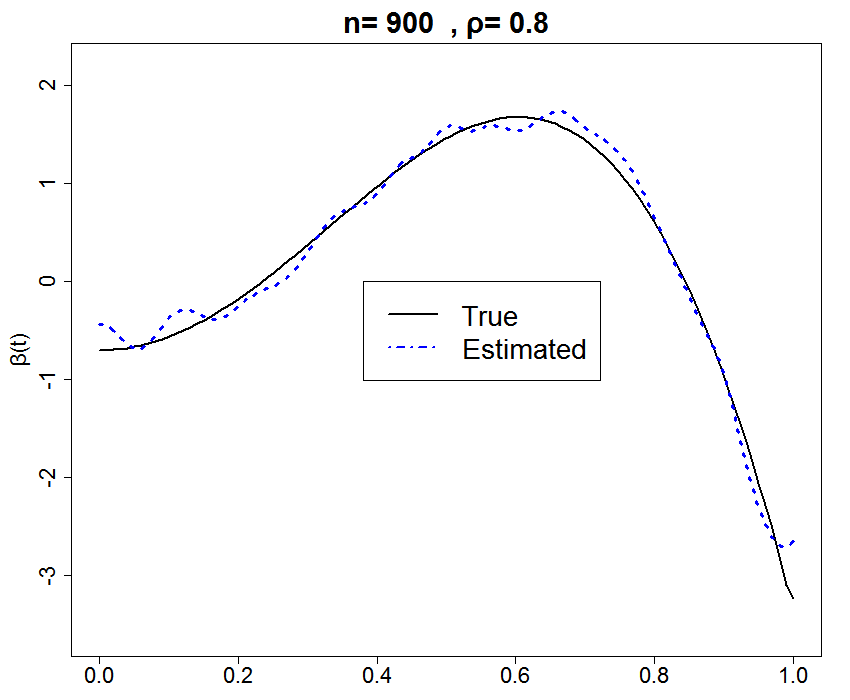}}}
  \end{center}
  \caption{The estimators of $\beta (t)$ vs the true $\beta (t)$ when the sample size $n=150, 300, 900$ and $\rho=0, 0.4, 0.8$. } \label{Fig.2}
\end{figure}

Table \ref{Tab 1} and Table \ref{Tab 2} report the performances of the estimators when $\alpha=1.1$ and $2$, respectively. We summarise the simulation results as follows.
\begin{enumerate}
  \item For $\hat{\rho}$, the mean biases are very close to $0$, and the standard deviations are small. When $\rho=0$, the biases of $\hat{\rho}$ can be positive or negative. When $\rho\neq0$, the biases are negative, which is similar to the results in \cite{Lee2004Asymptotic}.  In addition, the standard deviation has a decreasing tendency as the sample size increases. We also find that the variation in $\hat{\rho}$ decreases as $\rho$ increases.
  \item For the estimators $\hat{\beta}(t)$, $\hat{\beta}$ and $\hat{\bm{\beta}}^D$, their standard deviations (simplicial standard deviation) decrease as $n$ becomes larger as well. The mean biases of $\hat{\beta}$ and $\hat{\beta}^D$ are rather small. Note that summation of the mean biases of each part of $\hat{\beta}^D$ is $0$. We can also observe mean biases of $\beta(t)$ reduce rapidly with increasing sample size.
  \item All the parameters are estimated more accurately when $\alpha=1.1$ compared to $\alpha=2$. Specifically, the MSE of $\hat{\beta}(t)$ varies greatly when $\alpha$ takes on different values. \end{enumerate}

Figure \ref{Fig.2} displays the estimated slope function vs true coefficient function when $\alpha=1.1$. It can be observed that with increasing $n$, $\hat{\beta}(t)$ becomes closer to the true function.
In short, the simulation study demonstrates the efficiency of our proposed estimation method. The biases are small, and all estimators' standard deviations show a decreasing trend as the sample size increases. Moreover, the spaces between the eigenvalues of the sample covariance are crucial to the behaviours of the estimators.

\section{Real data analysis}\label{sec6}
In this section, we use the new model to analyse the factors affecting the PM2.5 concentrations, as mentioned in Section \ref{Sec1}.

\begin{figure}[!h]
  \begin{center}
    \resizebox{7cm}{5cm}{\includegraphics{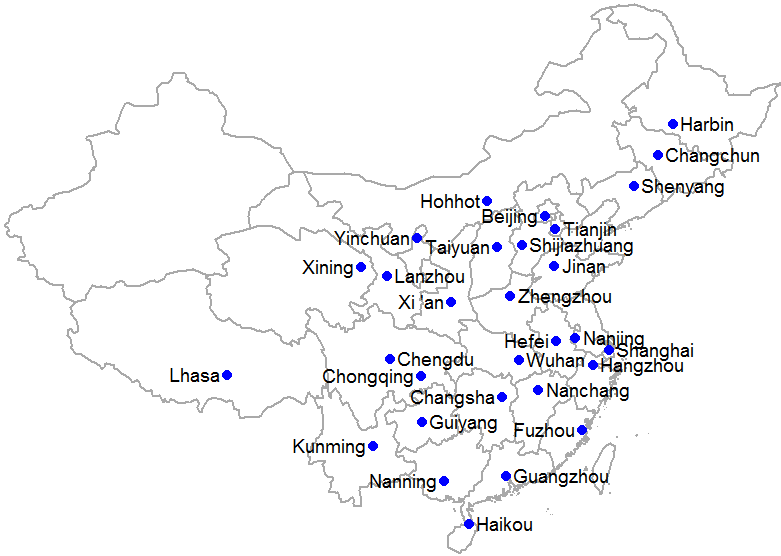}}
  \end{center}
  \caption{The locations of $30$ major cities.} \label{Fig.4}
\end{figure}

\begin{figure}[!h]
  \begin{center}
  \subfigure{
  \resizebox{4cm}{3.5cm}{\includegraphics{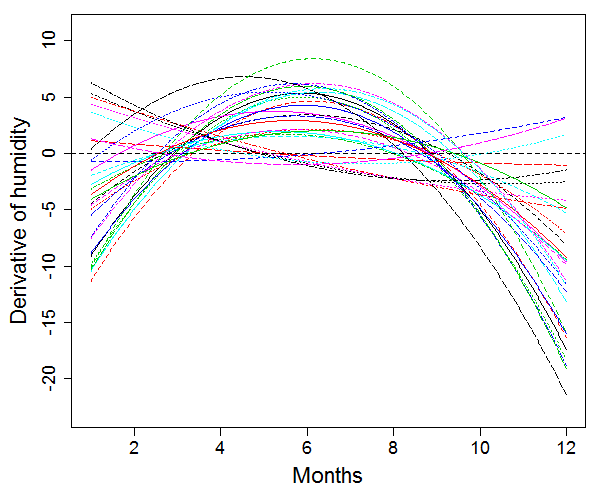}}}
   \subfigure{
  \resizebox{4cm}{3.5cm}{\includegraphics{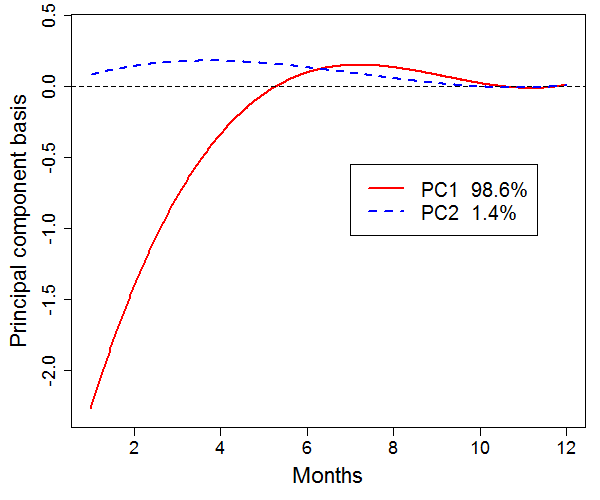}}}\\
  \subfigure{
  \resizebox{4cm}{3.5cm}{\includegraphics{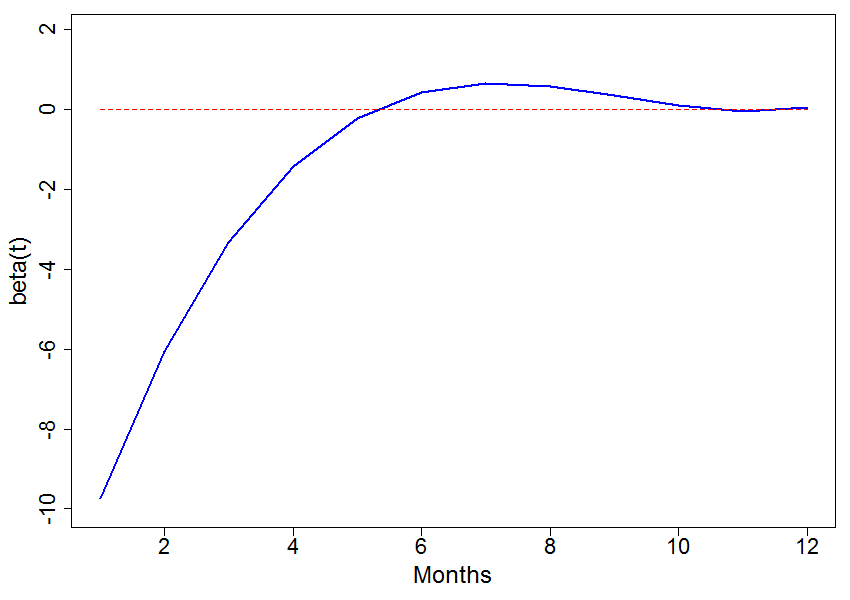}}}
  \subfigure{
  \resizebox{4cm}{3.5cm}{\includegraphics{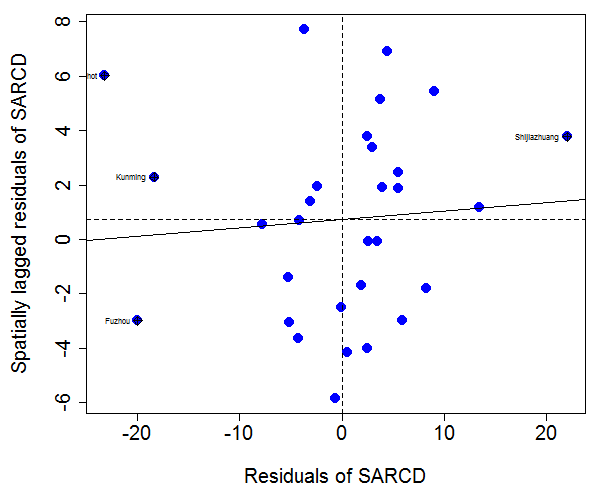}}}
  \end{center}
  \vspace{-3mm}
  \caption{The derivatives of the humidity curves in $30$ cities (top-left), principal component basis of the derivatives of the humidity curves (top-right), the estimated parameter function (bottom-left), and the Moran I scatter-plot of the residuals and the spatially lagged residuals of our model (bottom-right).} \label{Fig.5}
\end{figure}

Here, the PM2.5 concentration data,  humidity data, and economic data are collected from the China Air Quality Real-time Release Platform (http://106.37.208.233:20035), China Statistical Yearbook 2017 and Statistical Communique on the 2016 National Economic and Social Development, respectively. In the preprocessing of the data, the discrete recorded monthly relative humidity is transformed into continuous humidity curves by the Epanechnikov kernel. The output values of primary industry,  secondary industry and tertiary industry are transformed into compositional data, with each part representing the corresponding industry's percentage. Moreover, the derivatives of the humidity curves are evaluated. Figure \ref{Fig.5} (top-left) shows the derivatives of the functional humidities of $30$ cities.

The model for the PM2.5 data is formulated as follows:
\beq\label{real-equ}
y_i=\alpha+\sum_{i \neq i^{\prime}}\rho w_{ii^\prime}y_{i^\prime}+\int x_i(t)\beta(t)+\langle x_i^D,\beta^D\rangle+x_{i1}\beta_1+x_{i2}\beta_2+\epsilon_i,
\eeq
where $y_i$ is the annual mean PM2.5 concentration of the $i$th city, $x_i(t)$ is the $i$th city's  monthly humidity curve or derivative of the monthly humidity curve, $x_i^D$ is the composition of the three industries' percentages, $x_{i1}$ is the GDP growth rate, $x_{i2}$ is the logarithm of the GDP, and $w_{ii^\prime}$ is the weight between city $i$ and city $i^\prime$. The remaining $\alpha$, $\rho$, $\beta(t)$, $\beta^D$, $\beta_1$, and $\beta_2$ are the parameters to be estimated.

\begin{table}[!h]
 \begin{spacing}{2.0}
\footnotesize
\caption{ The values of the log-likelihood function of the $9$ alternatives of our model.}
  \setlength\tabcolsep{6pt}
  \renewcommand{\arraystretch}{1.5}
\label{Tab 3}
\bc
\begin{tabular}[H]{cccccccccccccccccccc}
\specialrule{0.05em}{-3pt}{-3pt}
 $\bm{k=2}$ & $\bm{k=3}$ & $\bm{k=4}$ & $\bm{k=5}$ & $\bm{k=6}$ & $\bm{k=7}$ & $\bm{k=8}$ & $\bm{k=9}$&$\bm{k=10}$ &\\
\specialrule{0.05em}{-3pt}{-3pt}
-113.68 & -114.21 & -112.22 & -110.10 & -110.64 & -112.41 & -112.43 & -112.41 &-112.20&\\ \specialrule{0.05em}{-3pt}{-3pt}
\end{tabular}\ec
\end{spacing}
\end{table}

We evaluate the $w_{ii^\prime}$ in model (\ref{real-equ}) according to the distance $d(i,i^\prime)$ between two cities $i$ and $i^\prime$. Specifically, $w_{ii^\prime}=\frac{1}{d(i,i^\prime)}$. Additionally, we assume that $w_{ii^\prime}=0$ if the distance $d(i,i^\prime)$ is greater than $15$. Figure \ref{Fig.4} presents the locations of our $30$ target cities on the map of China. Obviously, the weight between Haikou and Harbin is $0$ because they are far from each other. Here, we have many choices for the matrix $W=\{w_{ij}\}_{n \times n}$ regarding the maximum number of neighbours  $k$ . To choose an optimal weight matrix, we consider $9$ conditions: $k={2,3,4,5,6,7,8,9,10}$.

We note that the derivatives of the humidity curves achieve better fitting results than the humidity curves. Therefore, we use derivatives of the humidity functions as predictor in regression \ref{real-equ}.  Figure \ref{Fig.5} (top-right) shows the eigenfunctions of the FPCA. It can be observed that the first principal component (PC1) accounts for $98.6$ percent of the variation in the derivatives. Thus, PC1 approximates the predictor closely in this study. Also recall that there are $9$ alternatives for the weight matrix $W$. Therefore, considering number of principal components and choices of spatial weight matrix, a total of $9$ alternatives for the transformed model are considered here. Table \ref{Tab 3} summarises the values of the log-likelihood function under different $k$. Clearly, the weight matrix with $5$ nearest neighbours fits the model best. In this situation, we set $k$ to  $5$, and only the first principal component is involved in the parameter estimation.

\begin{table}[!h]
 \begin{spacing}{2.0}
\footnotesize
\caption{ The estimated parameters and their P values for our model.}
  \setlength\tabcolsep{6pt}
  \renewcommand{\arraystretch}{1.5}
\label{Tab 4}
\bc
\begin{tabular}[H]{cccccccccc}
\specialrule{0.05em}{-5pt}{-5pt}
 $ \bm{\hat{\alpha}}$ & $\bm{\hat{\beta}_1^D}$ & $\bm{\hat{\beta}_2^D}$& $\bm{\hat{\beta}_3^D}$&$\bm{\hat{b}_1}$&$\bm{\hat{\beta}_1}$&$\bm{\hat{\beta}_2}$&$\bm{\hat{\rho}}$\\ \specialrule{0.05em}{-5pt}{-5pt}
-21.21& $6.382858\times 10^{-4}$  & $9.993617\times 10^{-1}$ & $1\times 10^{-10}$&4.29&-1.04&6.82&0.62\\ \specialrule{0em}{-5pt}{-5pt}
 (0.27)& &  & &(0.019)&(0.08)&(0.002)&(0.0003)\\
   \specialrule{0.05em}{-5pt}{-5pt}
\end{tabular}\ec
\end{spacing}
\end{table}

Table \ref{Tab 4} is a summary of the estimated parameters of our model. First, we can see that the spatial dependencies are significance and play an important role in the regression. Second, GDP and  GDP growth are positively and negatively associated with PM2.5 concentration, respectively. Concerning relative humidity, we find that as the derivatives of the humidity curves increase during spring and summer, the PM2.5 concentrations decrease. Moreover, we find that the second part of the compositional coefficient accounts for a majority of the composition. Table \ref{Tab 5} displays the remaining results of our regression. We can see that our model eliminates most dependencies in the raw responses. Figure \ref{Fig.5} (bottom-right) demonstrates Moran's I scatter-plot of the residuals. In addition, the R-squared of the proposed model is 0.85, which means that most of the variances are explained by the network structure and predictors.

\begin{table}[!h]
 \begin{spacing}{2.0}
\footnotesize
\caption{ The fitting results of our model.}
  \setlength\tabcolsep{6pt}
  \renewcommand{\arraystretch}{1.5}
\label{Tab 5}
\bc
\begin{tabular}[H]{ccccc}
\specialrule{0.05em}{-5pt}{-5pt}
& R ~square& MSE~ of ~$y_i$ & $Moran's ~I ~statistic~ of~ residuals$ \\ \specialrule{0.05em}{-5pt}{-5pt}
& 0.85 & 82.09 & 0.03 (0.27) \\
   \specialrule{0.05em}{-5pt}{-5pt}
\end{tabular}\ec
\end{spacing}
\end{table}

\section{Conclusions}\label{sec7}
The mixed spatial autoregressive model fits data whose responses are dependent under a network structure and is very useful in spatial econometrics. Nevertheless, mixed SAR models only consider numerical covariates in regression, thereby being inflexible when complex data types are involved.

In this article, we consider functional, compositional and numerical predictors in an SAR model. Specifically, we mix these three types of data in a regression model by virtue of inner products defined in each data type's geometry space. The new model has the merits of a functional linear model, compositional linear model and MSAR model. In addition, we  present the estimators of the parameters of our model based on FPCA, the ilr transformation and MLE. During this procedure, we first transform the functional and compositional data into ordinary data, and then, we use MLE to obtain the estimators. Numerical experiments find that the standard deviations of the estimators show a decreasing trend when the sample size increases. In addition, regardless of the strength of the network influences, the parameters can be well estimated. Finally, we applied our method to a real PM2.5 dataset, which demonstrated the usefulness of our model.

It should be noted that our proposed model can be generalized. As long as the inner product for a new data type is defined, these new data can be added into our model. For example, interval data can also be contained in the new model. Under this setting, there are four mixed data types in the SAR model. These will be studied further in our next study.

\section*{Acknowledgements}

\noindent This research was financially supported by the National Natural Science Foundation
of China under grant nos. 71420107025 and 11701023.


\makeatletter
\addtolength{\@fpsep}{-12pt}
\makeatother

\makeatletter
\addtolength{\@fpsep}{-10pt}
\makeatother

\vspace{-10mm}

\bibliographystyle{smj}
\bibliography{SARCD-ref}
\end{document}